\documentclass{ws-p8-50x6-00}

\begin{document}

\title{Parton and dipole approaches in QCD}

\author{I.M. Dremin$^1$, P. Ed\'en$^2$}

\address{\it $^1$Lebedev Physical Institute, Moscow 119991, Russia\\
$^2$Nordita, Copenhagen, DK-2100, Denmark\\}

\maketitle

Here, we discuss QCD predictions on multiplicities in parton and dipole
approaches. The most general treatment is based on the notion of the
generating functions \cite{44}. The generating function $G$ is defined as
\begin{equation}
G(u,y)=\sum _nu^nP_n(y),
\end{equation}
where $P_n$ is the probability of the $n$-particle production at energy
denoted by $y$, $u$ is an auxiliary variable. The mean multiplicity and
higher moments of the multiplicity distribution $P_n$ are given by the
$u$-derivatives of $G$ at $u=1$.

The equations for the generating functions for gluon and quark jets have 
been proven in QCD up to the next-to-leading order (NLO) of the modified
perturbation theory \cite{5}. Their general structure is symbolically
represented as
\begin{equation}
G'\sim \int \alpha _SK[G\otimes G-G]d\Omega.       \label{symb}
\end{equation}
It shows that the evolution of $G$ indicated by its derivative
$G'$ over the evolution parameter (the transverse momentum or the virtuality)
 is determined by the cascade process of the production of two partons
by a highly virtual time-like parton (the term $G\otimes G$) which provides
new partons in the phase space volume $d\Omega $ and by the
escape of a single parton ($G$) from a given phase space region.

Therefore
this equation contains terms corresponding to inflow and outflow of partons.
In fact, it can be interpreted as the kinetic equation with the collision
integral in the right hand side. The weight factors are determined by the
coupling strength $\alpha _S$ and the splitting function $K$ which is defined
by the interaction Lagrangian. The integral runs over all internal
variables, and the symbol $\otimes $ shows that the two created partons 
share the
momentum of their parent. The initial condition for equation (\ref{symb})
is defined by the requirement for the jet to be created by a single initial
parton, i.e., by 
\begin{equation}
P_n=\delta _{n1}; \;\;\;\;\;\;\;\;\; G_0=u.
\end{equation}
It is clear from this formula that we have to deal with the non-linear
integro-differential probabilistic equation with shifted arguments in
the $G\otimes G$ term under the integral sign.

For the traditional parton approach \cite{5}, the energy conservation
at cascade vertices is properly accounted by the shift of corresponding
arguments of the generating functions in the integral whereas the transverse
momentum ($p_t$) limitations are considered approximately as a combined 
effect of energy conservation and angular ordering.

In the dipole QCD evolution advocated by the Lund group \cite{ande},
the triangle phase space in the energy - transverse momentum plane is
considered. Limitations on the available phase space (both on energy and
$p_t$) due to recoil effects, emission of additional gluons, high-$p_t$
processes, color reconnection etc can be explicitly implemented \cite{egus}.

Both approaches agree up to NLO of the perturbative QCD (pQCD) \cite{gust}.
However, the corresponding equations for the generating functions differ
at higher orders \cite{eden}.

Some other modifications of these equations have been earlier proposed
\cite{123, 127}. Thus, the problem of formulation of general equations
for generating functions can not be still considered as solved. On the
way to its solution, we tried to understand how strong is the difference
between the parton and dipole approaches in this respect. The partial
answer is presented in this talk.

First, let us describe the present situation up to NLO where both approaches
coincide. The analytical solutions of the equations successfully predict
the energy behavior of mean multiplicities \cite{5}. However, the theoretical
values of the ratio $r$ of multiplicities between gluon and quark jets are
larger than experimental data (at Z$^0$, by about 50$\%$ for the leading order
(LO) which gives asymptotic values and by 30$\%$ for NLO). The description
of higher moments is also not perfect \cite{dlne}. However, in each case
NLO corrections improve the agreement compared with LO results. Moreover,
pQCD has predicted \cite{13} in NLO the new unexpected feature of the
behavior of cumulant moments which become negative at higher ranks. Their
ratio to factorial moments ($H_q$) as a function of $q$ acquires the 
minimum at $q\approx 5$. It has been confirmed by experiment. Many other
features of multiparticle production have also been explained in NLO of
pQCD \cite{kowo}.

To go beyond NLO, one of the strategies is to consider the parton 
evolution equation as a kinetic equation and find its perturbative 
solution. The systematic method of the Taylor series expansion \cite{13}
leads to the perturbative series for the ratio $r$ and the anomalous
dimension $\gamma $ which determines the energy behavior of mean 
multiplicity.
In such a way, the results up to 3NLO have been obtained \cite{cdnt8}.
High order corrections almost do not influence conclusions about the energy
dependence of mean multiplicity. However, they improve the agreement with 
experiment on the ratio $r$ (up to 15$\%$). High order terms completely
determine the energy dependence of $r$ (its slope) \cite{drsl} because 
the main (NLO) dependence is the same for gluon and quark jets and it 
cancels
in their ratio. Moreover, the oscillations of $H_q$-moments with $q$ have
been predicted both for running \cite{41} and fixed \cite{dhwa} coupling
regimes at high orders and found in experiment (first in \cite{dabg} and then
in \cite{sld}). However, the direct perturbative calculations of moments
showed that the high order results become unreliable. Formally, this is due
to the fact that the expansion parameter $\gamma _0\propto \alpha _S^{1/2}$,
where $\alpha _S$ is the QCD coupling strength, becomes multiplied by the 
rank $q$, and this product is larger than 1. In practice, this means that
soft low-$p_t$ partons become important. While purely perturbative methods
fail here, the exact solution for fixed coupling \cite{dhwa} and numerical
computer solutions \cite{lo} show much better results. In more detail, the
parton approach is reviewed in \cite{dgar}.

To cure these problems, one can try another strategy and modify the parton
cascade equations so that NLO results for new and old versions coincide.
Such a modification
inspired by the dipole model has been proposed \cite{eden}. It noticeably
improves the agreement in 2NLO order with experimental data on $r$ at 
Z$^0$.
The agreement results from different (from parton approach) boundary 
condition
imposed with account of $p_t$-conservation. A detailed comparison of dipole
and parton formalism~\cite{eden} reveals that sensitivity to cascade choice,
and hence infrared cutoff quantity representing hadronization details, give
uncertainties to 2NLO which reduce the value of higher order
refinements on $r$.

However, previous experience with parton equations has taught us that higher
moments are very sensitive to subtle modifications due to phase space
limitations and hadronization effects. Our recent results support this
statement. To be able to treat the dipole equations analytically, we use the
perturbative expansion. Matching the terms of the same order on both sides
of the equations, we get the moments up to (in principle, any) predefined 
order. The dipole evolution expanded to high orders does not
converge towards oscillating $H_q$-moments. Instead, higher rank moments
diverge severely, and the only hint of oscillations lies in the sign
change of divergence for each new order added in the expansion as seen
in Fig. 1 for $H_q$ at Z$^0$. 

The similar sign-changing curves one obtains if the Taylor series of any
oscillating function (e.g., sine ) is cut off at some high order terms.
Let us note that in previous work on parton equations \cite{41}, which
revealed explicit oscillations, some kind of Pade approximation was used
where terms up to a definite order were kept both in the numerator 
(cumulants)
and denominator (factorial moments) of $H_q$. This is plausible in view of
large expansion parameter at hand.

>From this discussion we conclude that the slope of the ratio $r$ and high 
rank 
moments of multiplicity distributions are most sensitive to generalization of 
master evolution equations to higher orders of the perturbative expansion.
 
For this improvement to be successful, one must solve several problems. 
First,
the calculations apply to parton multiplicities which then are identified with
hadron data using the assumption of the local parton-hadron duality (LPHD)
\cite{adkt}.
We must also note that pure perturbative QCD predictions can be obtained 
only
for infrared safe quantities. For multiplicities, one has to deal with some
cut-off parameter specifying the boundary condition which is influenced by
the non-perturbative region of soft partons and hadronization scheme.
In general, the cut-off is determined locally in phase space, related to
a matching of colors and anti-colors into small singlet sub-systems, often
referred to as the 'preconfinement' assumption. At low scales, the
non-perturbative boundary conditions are essential and the energy scale 
where an asymptotic expansion of moments becomes useful depends on
assumptions about hadronization. In the
simplest approximation, "extreme LPHD" if you like, each parton is
traded for one hadron at some cut-off scale, usually around 2 to 5
$\Lambda_{QCD}$. How such a boundary condition at a low scale
influences the approach to the asymptotic result can be investigated
with numerical integration on computer. One such result, $K_2$ in
gluon dynamics, is presented in Fig. 2. We note that the
differences implied by varying the boundary scale from 2 to 4
$\Lambda_{QCD}$ prevails beyond Z$^0$ energies.

One could of course imagine less extreme hadronization schemes, e.g.,
letting each parton represent a small distribution of hadrons, with
some low average. In the figure is also shown the result assuming a
poissonian with unit average. The difference is drastic, but it should
be admitted that the poissonian has a rather unrealistic tail to large
multiplicities. Note however, that factorial moments obtained with a
poissonian boundary condition are identical to 'normal' moments starting
with delta-distribution (extreme LPHD). Thus the figure could serve as 
inspiration to a simultaneous experimental study of both factorial and
normal moments, which may give additional insight to multiplicity boundary
conditions.

In conclusion, analytical treatment of the iterative cascade has revealed
energy conservation and the "preconfinement" hadronization assumption as 
most
significant for results on $r$ at available energies. Whether the
analytical approach can be of similar use to understand in more detail
higher multiplicity moments remains to be seen. Apparent is that
multiplicity distributions are very powerful probes of the
perturbative - nonperturbative transition in QCD. We hope that our
contribution will give somebody a hint to further work on generalization
of QCD master evolution equations.

\end{document}